\documentclass[runningheads]{llncs}
\usepackage[T1]{fontenc}
\usepackage{graphicx}
\usepackage{color}
\usepackage{booktabs}
\usepackage{multirow}
\usepackage{cite}
\begin{document}
%
\title{\Large \bf A Unified Evaluation of Learning-Based Similarity Techniques for Malware Detection }
%
%

\author{
{\rm Udbhav Prasad}\inst{1}\\
{\rm Aniesh Chawla}\inst{1}\\
} 
%
%
\institute{Independent Researchers, California, USA \\
\email{\{udbhav523, chawla.aniesh\}@gmail.com}
\url{https://anonymous.4open.science/r/malware-clustering}}
\maketitle              
%

%
\begin{abstract}
Cryptographic digests (e.g., MD5, SHA-256) are designed to provide exact identity. Any single-bit change in the input produces a completely different hash, which is ideal for integrity verification but limits their usefulness in many real-world tasks like threat hunting, malware analysis and digital forensics, where adversaries routinely introduce minor transformations. Similarity-based techniques address this limitation by enabling approximate matching, allowing related byte sequences to produce measurably similar fingerprints. Modern enterprises manage tens of thousands of endpoints with billions of files, making the effectiveness and scalability of the proposed techniques more important than ever in security applications. Security researchers have proposed a range of approaches, including similarity digests and locality-sensitive hashes (e.g., ssdeep, sdhash, TLSH), as well as more recent machine-learning–based methods that generate embeddings from file features. However, these techniques have largely been evaluated in isolation, using disparate datasets and evaluation criteria. This paper presents a systematic comparison of learning-based classification and similarity methods using large, publicly available datasets. We evaluate each method under a unified experimental framework with industry-accepted metrics. To our knowledge, this is the first reproducible study to benchmark these diverse learning-based similarity techniques side by side for real-world security workloads. Our results show that no single approach performs well across all dimensions; instead, each exhibits distinct trade-offs, indicating that effective malware analysis and threat-hunting platforms must combine complementary classification and similarity techniques rather than rely on a single method.
\end{abstract}
\section{Introduction}
Modern malware analysis systems operate at massive scale, routinely processing billions of executable files across enterprise environments. In this setting, malware triage extends beyond binary (malware / benign) detection to encompass tasks such as family attribution and clustering, which depend on the ability to identify similar binaries. Malware authors apply packing, recompilation, instruction reordering, and minor structural changes that render exact cryptographic hashes (e.g., MD5, SHA-256) ineffective for these downstream analysis tasks. As a result, approximate matching and similarity-based fingerprinting have become foundational capabilities in practical malware analysis pipelines.

This challenge is commonly formalized as Binary Code Similarity (BCS), which seeks to quantify the similarity between compiled binaries in the absence of source code and under diverse compilation and obfuscation conditions.

To achieve these goals, researchers and industrialists leverage two distinct methodologies:  dynamic analysis and static analysis. Dynamic analysis involves running the binary in a sandboxed environment to observe runtime behavior and identify any potential harm. Conversely, static analysis examines the binary's properties - compiler name, file headers, byte entropy, etc. This paper focuses on the latter and uses the EMBER~\cite{emberdataset} dataset for precomputed static analysis of 1 million files. New techniques ~\cite{chawla2026decompilation} have also been proposed in which the binary code is decompiled into source code and language models are used for malware detection.

Recent advances in machine learning have enabled a shift toward feature-based and learning-driven similarity representations that derive dense embeddings from static PE metadata, including headers, imports, and entropy-based features. Neural and tree-based models offer the potential to capture higher-level structural relationships while supporting efficient indexing and nearest-neighbor retrieval. However, existing studies typically evaluate these approaches in isolation, often emphasizing classification accuracy rather than the quality of the induced similarity space. As a result, it remains unclear how classical fuzzy hashes and heterogeneous learning-based embeddings compare under a unified similarity framework, or what trade-offs they present for real-world malware triage and threat-hunting workloads.

In this paper, we present a systematic and reproducible comparison of classical similarity digests and learning-based embedding techniques for malware analysis under a unified experimental framework. Using large-scale static PE metadata from the EMBER and EmberSim \cite{crwdstrkepaper} datasets, we evaluate fuzzy hashing alongside embeddings derived from unsupervised autoencoders, deep neural network classifiers, and gradient-boosted decision trees. All methods are assessed using consistent distance functions and industry-accepted metrics for binary detection, multi-class malware family classification, and similarity evaluation. Our study provides the first side-by-side, metric-consistent benchmarking of these heterogeneous similarity representations on a shared dataset, revealing that no single approach dominates across all dimensions and that effective malware analysis systems must integrate complementary similarity techniques rather than rely on a single method.
\section{Related Work}

\subsection{Binary Code Similarity and Conventional Hashing}
Binary Code Similarity (BCS) is the quantitative comparison of two binary artifacts to determine their functional or structural similarity \cite{haq2019surveybinarycodesimilarity}. Early research in static-based BCS primarily focused on specific components, such as function call graphs \cite{malwareindexing, similarityFunctionCall} or isolated static features \cite{mutantXclustering}, rather than a complete, full program analysis. Thein et al.~\cite{thein2025comparativeanalysishashbasedmalware} explored the fuzzy hashing techniques for the K-means clustering. However, the authors didn't explore machine learning based clustering techniques.

In industrial settings, fuzzy hashing remains the standard for rapid similarity assessment. Algorithms such as ssdeep~\cite{ssdeep}, sdhash~\cite{sdhash}, and TLSH~\cite{tlsh} generate similarity scores by processing byte-level blocks within the input. While computationally efficient, these byte-centric schemes are often brittle when faced with minor structural permutations or compiler-induced variances. They also don't have the advantage of machine learning in learning from enterprise-specific patterns for better detection accuracy.

\subsection{Machine Learning and Feature-Based Approaches}
Recent years have seen a transition toward learning-based similarity models. A notable contribution in this domain is EmberSim \cite{crwdstrkepaper}, that utilizes an XGBoost model where leaf node values serve as embeddings to identify similarity among nearby binary files. However, a significant limitation of this approach is its reliance on the avclass index during training. Because avclass labels are exclusive to malicious samples, this methodology introduces a classification bias as shown in \ref{subsect:label_homogeneity}, potentially skewing the model’s ability to generalize similarity across benign datasets. Priyanka et al.~\cite{9750242} and Bamidele et al/~\cite{ajayi2024exploring} leverages the transfer learning technique for malware classification but do not show malware clustering analysis on these.

Other approaches, such as MalMixer \cite{malmixer}, employ semi-supervised learning to generate synthetic samples. This method requires significant manual intervention to distinguish between "interpolatable" and "non-interpolatable" features. Furthermore, the synthetic samples generated are not guaranteed to be functional or even valid when reverse engineered. Feature importance has also been a focus of research; for instance, Oyama et al. \cite{8951564} identified critical features within the EMBER dataset for malware classification but relied on supervised methods rather than allowing an unsupervised model to discover latent feature sets. More recently, Abedin et al. \cite{abedin2025} performed a comparative analysis of LightGBM, XGBoost, CatBoost, and TabNet on the EMBER dataset using dimensionality reduction techniques like PCA and LDA.

\subsection{LLMs and Decompilation-Based Analysis}
The state-of-the-art has recently shifted toward the use of Large Language Models (LLMs). Some approaches involve passing decompiled binary code directly into LLMs for analysis \cite{chawla2026decompilation}. While these models show promise in understanding high-level logic, they frequently lack explainability. Furthermore, the token limits inherent in current LLM architectures often force researchers to discard significant portions of the file's static metadata, potentially losing rich contextual information that non-textual static analysis would otherwise capture.

\subsection{Our Contribution}
In our paper, we address a lack of unified evaluation for heterogeneous similarity techniques in malware analysis. Our primary contributions are as follows:

\begin{itemize}
    \item \textbf{Unified Benchmarking Framework:} We present a reproducible study to benchmark classical similarity digests (e.g., ssdeep), side by side with learning-based embeddings using industry-accepted metrics.
    \item \textbf{Systematic Feature Engineering for PE Metadata:} We develop a rigorous pre processing pipeline for the EMBER dataset that transforms hierarchical JSONL metadata into fixed-length numerical vectors through structural normalization, schema flattening, and vectorization.
    \item \textbf{Comparative Analysis of Learning-Based Embeddings:} We evaluate and compare three distinct machine learning approaches to derive latent representations: unsupervised Autoencoders (AE), binary deep learning classifiers, and tree-based XGBoost leaf-index embeddings.
    \item \textbf{Empirical Performance Validation:} Our results demonstrate that learning-based approaches, specifically unsupervised Autoencoders, significantly outperform traditional fuzzy hashing techniques like ssdeep, achieving over 80\% label homogeneity in Top-K similarity queries compared to approximately 40\% for classical methods.
    \item \textbf{Malware and AVClass Classification:} We provide a comprehensive performance analysis of deep learning and gradient-boosted models for both binary detection and multi-class AVClass family classification, achieving up to 97.76\% accuracy in binary tasks.
\end{itemize}
\section{Data}
EMBER~\cite{emberdataset} is an open dataset of 1M samples containing static PE details for binary files that were scanned before 2018. The data consists of 800k labelled samples out of which 400k are malware and 400k are benign. The remaining 200k samples are unlabeled.
It has been used to train various machine learning models for malware identification~\cite{vinayakumar2019robust, 10068497, 10681072, pham, wuEnhancing} and malware classification ~\cite{crwdstrkepaper, 9750242, ajayi2024exploring}. The EmberSim~\cite{crwdstrkepaper} dataset further enhances the 2018 Ember dataset with another added \textit{AVClass}. They added these feature based on \textit{Tag enrichment via co-occurrence} technique. We have used this dataset in our experiments.

\subsection{EMBER Features}

A PE file \cite{microsoft_pe_format} is a structured Windows executable containing headers, sections, and metadata such as imports, resources, and relocation information. The EMBER dataset uses the LIEF \cite{LIEF} tool to extract PE metadata into JSONL format (newline-separated series of JSON records). The metadata included in the EMBER dataset covers different data types:
\begin{itemize}
    \item Basic data types: scalar values including file hashes and header fields
    \item Array types: variable-length feature sets such as imports, exports, and section-level attributes
    \item Structured types: hierarchical representations of PE headers and directory metadata
\end{itemize}

\subsection{Feature Engineering}

The raw metadata extracted from PE files consists of nested and heterogeneous structures and is not directly suitable for consumption by standard machine learning models, which require fixed-length numerical feature vectors. Our feature engineering process preserves discriminative feature information through a combination of normalization, dictionary hashing, and aggregation. The process is divided into two primary stages: structural normalization and numerical vectorization.

\subsubsection{Structural Normalization and Extraction}
First non-informative identifier fields (e.g., cryptographic hashes) are excluded. Then the initial stage focuses on transforming the hierarchical JSONL metadata into a structured, flat schema. We define a rigorous schema to parse the raw metadata, ensuring that all records adhere to a consistent format. 

\begin{itemize}
    \item \textbf{Schema Flattening:} Nested structures within the PE headers such as COFF headers and data directories are flattened into discrete columns. For example, the nested \texttt{imports} object, which maps libraries to specific functions, is unrolled into a flat array of imported function strings.
    \item \textbf{Array Processing:} Variable-length arrays including \texttt{exports} and \texttt{sections} are extracted into unified string lists. For numeric arrays that represent frequency distributions, such as the byte \texttt{histogram}, \texttt{byteentropy}, and \texttt{printabledist}, we convert the raw values into high-density numeric vectors.
    \item \textbf{Data Cleaning:} To ensure experimental integrity, we apply a filtering pass to remove any records with malformed parsing results and deduplicate the dataset based on the unique \texttt{sha256} file hash.
\end{itemize}

\subsubsection{Numerical Vectorization}
Once the data is flattened, we apply a series of transformations to convert categorical and list-based attributes into fixed-length numerical inputs required by machine learning architectures.

\begin{itemize}
    \item \textbf{Categorical Encoding:} For fixed-category string fields such as the \texttt{avclass} labels, we employ string indexing to map each unique category to a unique numeric identifier.
    \item \textbf{Bag-of-Words Vectorization:} To handle the high-cardinality and variable-length nature of imports, exports, and section characteristics, we use a count vectorization approach. We establish a fixed vocabulary limit of 2,048 for each feature set, mapping the occurrence of specific attributes into a binary or frequency-based feature vector.
    \item \textbf{Dimensionality Finalization:} After the vectorization is complete, the original high-dimensional string columns and intermediate indices are discarded, resulting in a finalized feature set where every attribute is represented as a scalar or a fixed-length vector.
\end{itemize}
\section{Methodology}

We train binary classifier models against the malware / benign \textit{label}s in the original EMBER~\cite{emberdataset} dataset, and additionally train multi-class classifier models against the \textit{AVClass} tags added in the EmberSim~\cite{crwdstrkepaper} dataset.

We further use the trained models to derive dense feature embeddings. They act as "feature extractors" that compress high-dimensional PE data into a latent representation of the input sample.

\subsection{Autoencoder}
\label{subsect:autoencoder}

The Autoencoder (AE) is an unsupervised neural network architecture designed for efficient dimensionality reduction and feature extraction. By compressing high-dimensional input data into a lower-dimensional representation, the AE facilitates the capture of salient features while filtering out stochastic noise. This makes it a robust pre-processing step for Ember and EmberSim feature sets. In this study, the AE architecture utilizes a symmetric multilayer perceptron structure with hidden layers of sizes [256, 64, 8, 64, 256]. Given the high cardinality of the initial feature set, a relatively wide architecture was selected to preserve structural information during the initial compression phases. The central bottleneck, or latent space, consists of 8 dimensions, which we hypothesized to be sufficient for representing the original 256-feature input space without significant loss of critical information.The model was trained using the Adam optimizer, with Mean Squared Error (MSE) serving as the objective function to minimize reconstruction loss. The relationship between the Encoder ($E$) and the Decoder ($D$) is formally defined by the loss function $\mathcal{L}$, which measures the divergence between the input $x$ and the reconstructed output $x'$ :
\begin{equation}
\mathcal{L}(\theta, \phi) = \frac{1}{M} \sum_{j=1}^{M} \|x^{(j)} - D_{\phi}(E_{\theta}(x^{(j)}))\|^2    
\end{equation}

In our experimental setup, a 25\% subset of the labeled data was utilized to train the aforementioned autoencoder model. Once the model reached convergence, it served as an encoder to generate the embeddings for the rest of the data points thereby reducing the dimensionality of the dataset. The resulting embeddings are then used to create a \textbf{Vector Index}. This is done to optimize the retrieval performance for high computational queries required to identify the Top-K similar items.

\subsection{Binary Classifier -- Deep Learning}

\subsubsection{Train-test split}
The \textit{label} denotes whether the sample is malicious or benign, and is used as the target for the classifier. Samples with \textit{Unknown} label values are ignored, resulting in a total of 800,000 samples in the final dataset, with an equal distribution of 400,000 malicious and benign samples. The dataset is partitioned into training and test sets using a stratified split to preserve the original class distribution. To ensure numerical stability and accelerate convergence, all vectorized features are standardized using z-score normalization computed exclusively on the training split, with the same transformation applied to the test split to avoid data leakage.

\subsubsection{Network architecture}
We employ a fully connected deep neural network consisting of the following components:
\begin{itemize}
    \item A sequence of dense layers with decreasing width: 512 → 256 → 128 units
    \item Each dense layer is followed by batch normalization, ReLU activation, and dropout for regularization
    \item A dedicated embedding layer of dimension 128, which serves as a compact latent representation of each sample
    \item A final sigmoid-activated output layer producing a malware probability score
\end{itemize}

Dropout with a rate of 0.3 is applied to mitigate overfitting, while batch normalization stabilizes training across feature groups with heterogeneous scales. The model is trained using the Adam optimizer with a learning rate of 0.001, binary cross-entropy loss and early stopping. In addition to accuracy, we monitor AUC, precision, and recall throughout training to better capture performance under class imbalance.

\subsubsection{Embedding extraction}
This classifier serves and as a foundation for learning semantically meaningful representations of executable files. We define a secondary model that maps inputs to the activations of the 128-dimensional embedding layer. Because the embedding layer is trained jointly with the classification objective, it captures discriminative features relevant to malware detection. This approach allows us to decouple representation learning from the classification head while preserving task-specific semantics.

\subsection{AVClass Classifier -- Deep Learning}
\label{subsect:avclass_dl}

\subsubsection{Filtering top-N AVClass values}
The \textit{AVClass} label space in the EMBER 2018 dataset exhibits a highly skewed long-tailed distribution, with a substantial majority of families represented by only a small number of samples. The dataset has 3226 distinct \textit{AVClass} values. 3038 (94.1\%) of the \textit{AVClass} values have under 200 distinct samples, 2940 (91.1\%) \textit{AVClass} have under 100 distinct samples and 2377 (73.6\%) \textit{AVClass} values have lower than 10 distinct samples. Classes with few distinct samples provide no meaningful signal for supervised learning. The model will tend to overfit individual samples and fail to generalize. It may also lead to spurious correlations unrelated to family semantics. Evaluation metrics like macro-F1 score become dominated by noise.

To mitigate this issue, we restrict multiclass classification experiments to the top 200 \textit{AVClass} labels by frequency, which together account for the majority of labeled malware samples and provide adequate representation per class. This filtering yields a well-defined and balanced learning problem while preserving meaningful family-level structure.

\subsubsection{Train-test split}

As in the binary classification setup, identifier fields are removed and all numeric features are standardized using z-score normalization computed on the training set and applied consistently across validation and test splits. We also apply class weighting during optimization based on inverse class frequency.

\subsubsection{Network architecture}

We employ a deep fully connected neural network whose architecture closely mirrors that of the binary deep learning classifier, enabling controlled comparison between binary and multiclass objectives. The main difference is the classification head, which is a dense layer with 200 units followed by a softmax activation, corresponding to the top-200 AVClass labels. The multiclass model is trained using the categorical cross-entropy loss and the Adam optimizer with a learning rate of 0.001. During training, we monitor overall accuracy as well as macro-averaged precision and recall to capture performance across both frequent and less frequent families.

\subsection{Binary Classifier -- XGBoost}
\label{subsect:binary_xgboost}

\subsubsection{Train-test split}
As in the neural network–based binary classifier, identifier fields are excluded and all numeric features are standardized using statistics computed on the training set only. The dataset is partitioned into training and test subsets using a stratified split over the binary labels, preserving the original benign-to-malware ratio. This preprocessing ensures consistency across models and enables a fair comparison between tree-based and neural approaches.

\subsubsection{XGBoost classifier}
We employ a gradient-boosted decision tree ensemble for binary malware classification. Tree-based models are well suited to EMBER-style feature representations due to their ability to model non-linear feature interactions and naturally handle heterogeneous feature groups without requiring explicit feature transformation.

The model is trained using a logistic loss objective, producing probabilistic predictions of maliciousness. Key hyperparameters governing model capacity and optimization behavior include tree depth, number of boosting iterations, and learning rate. These hyperparameters are selected via cross-validated grid search over a restricted set of candidate configurations. Models are evaluated using classification performance on held-out folds, and the configuration with the best average performance is selected. This reduced yet principled tuning strategy enables robust hyperparameter selection without incurring prohibitive computational cost.

\subsubsection{Embedding extraction}
The trained gradient-boosted decision tree model is repurposed as an embedding extractor by leveraging its internal tree structure. For each input sample, we record the terminal leaf reached in each tree of the ensemble. Because the model consists of 100 trees, this procedure yields a 100-dimensional embedding vector, where each dimension corresponds to the index of the leaf node reached in a particular tree.

These leaf-index representations can be interpreted as embeddings in which samples that traverse similar decision paths through the ensemble are close in the induced feature space. Prior work has shown that tree-based proximity and leaf-index representations preserve both geometric structure and classification-relevant information, making them well suited for downstream similarity analysis and clustering tasks \cite{rhodes2023geometry}.

\subsection{AVClass Classifier -- XGBoost}
We train a multi-class XGBoost model for AVClass family classification using the same top-200 label filtering strategy described in Section \ref{subsect:avclass_dl}. The dataset preprocessing pipeline, feature normalization, and stratified train-test split follow the procedure outlined for the deep learning classifier.

The model architecture and hyperparameter tuning methodology mirror the binary XGBoost configuration described in Section \ref{subsect:binary_xgboost}. The primary difference is the use of the \textit{multi:softprob} objective to optimize multi-class log-loss over the 200 AVClass families.

As with the binary XGBoost model, the trained ensemble consists of 100 trees. Each sample is represented by the indices of the terminal leaves reached across all trees, yielding a 100-dimensional leaf-index embedding. Embedding extraction is performed identically to the method described in Section \ref{subsect:binary_xgboost}.

Because this model is trained directly on malware family labels, the resulting embedding space is explicitly optimized for inter-family separation. These embeddings are subsequently evaluated under the unified similarity framework described in Section \ref{sec:evaluation}.
\section{Evaluation}
\label{sec:evaluation}

\subsection{Binary (Malware) Classifiers}

Binary classification performance is evaluated on a held-out test set using standard metrics -- accuracy, precision, recall, and area under the receiver operating characteristic curve (AUC). In our case, since there is no class imbalance in the binary classification case, AUC and recall do not provide much additional information about the model’s ability to distinguish malicious from benign samples across decision thresholds.

\subsection{Multi-class (AVClass) Classifiers}

For malware family classification, evaluation is performed over the top-200 AVClass labels on a held-out test set. Because this task involves moderate class imbalance and a large number of classes, we report both overall accuracy and macro-averaged precision and recall. Macro-averaged metrics weight each class equally and therefore provide a more informative assessment of performance across both frequent and less frequent families. These metrics capture the model’s ability to discriminate between prevalent malware families using static features alone, while avoiding overemphasis on dominant classes.

\subsection{Similarity}

To evaluate similarity, we assess the quality of the learned embedding spaces using three widely adopted internal clustering metrics: the \emph{Silhouette Score}~\cite{rousseeuw1987silhouettes}, the \emph{Davies--Bouldin Index}~\cite{davies1979cluster}, and the \emph{Calinski--Harabasz Index}~\cite{calinski1974dendrite}. These metrics provide complementary perspectives on cluster cohesion and separation without requiring external supervision beyond class labels.

All similarity and clustering metrics are computed using the \textbf{Euclidean distance (L2 norm)} between embedding vectors. Prior work such as EmberSim~\cite{crwdstrkepaper} evaluates similarity using \emph{tree proximity--based distances}, where similarity is derived from the fraction of shared leaf nodes across trees in a gradient-boosted ensemble. While effective for tree-based models, this distance measure is inherently tied to the internal structure of decision trees and does not naturally extend to embeddings produced by neural networks or other continuous representation learners. To ensure methodological consistency and fair comparison, we require a \emph{single distance function applicable across all embedding types}. Euclidean distance satisfies this requirement, as it operates directly on numeric vectors without relying on model-specific semantics.

\subsubsection{Silhouette Score}
The Silhouette Score measures how well each sample lies within its assigned cluster relative to other clusters. For a given sample $i$, let $a(i)$ denote the average Euclidean distance to all other samples in the same cluster, and let $b(i)$ denote the minimum average Euclidean distance to samples in any other cluster. The Silhouette Score $s(i)$ is defined as:
\begin{equation}
s(i) = \frac{b(i) - a(i)}{\max(a(i), b(i))}.
\end{equation}

The score ranges from $-1$ to $1$, where higher values indicate better-defined and more clearly separated clusters. Silhouette analysis captures per-sample cluster assignment quality and is particularly informative for evaluating embedding spaces used in nearest-neighbor similarity search.

\subsubsection{Davies--Bouldin Index}
The Davies--Bouldin Index (DBI) evaluates clustering quality by measuring the average similarity between each cluster and its most similar neighboring cluster. Using Euclidean distance, the DBI is defined as:
\begin{equation}
\mathrm{DBI} = \frac{1}{|C|} \sum_{i=1}^{|C|} \max_{j \neq i}
\left(
\frac{S_i + S_j}{M_{ij}}
\right),
\end{equation}
where $S_i$ denotes the average intra-cluster Euclidean distance for cluster $i$, and $M_{ij}$ is the Euclidean distance between the centroids of clusters $i$ and $j$. Lower DBI values indicate tighter clusters and greater inter-cluster separation.

\subsubsection{Calinski--Harabasz Index}
The Calinski--Harabasz (CH) Index measures the ratio of between-cluster dispersion to within-cluster dispersion:
\begin{equation}
\mathrm{CH} =
\frac{\mathrm{Tr}(B_k) / (k - 1)}
     {\mathrm{Tr}(W_k) / (n - k)},
\end{equation}
where $\mathrm{Tr}(B_k)$ and $\mathrm{Tr}(W_k)$ are the traces of the between-cluster and within-cluster scatter matrices computed using Euclidean distances, $k$ is the number of clusters, and $n$ is the total number of samples. Higher CH values indicate stronger global cluster separation and lower intra-cluster variance.

\subsubsection{Label Homogeneity}
The degree to which items within a cluster or group share the same label or class. In the context of binary code similarity, it measures how consistently similar binaries are grouped according to their underlying characteristics, such as originating from the same source code, compiler, or functionality. A high label homogeneity indicates that binaries within a cluster mostly share the same label, suggesting that the similarity metric or embedding captures meaningful structural or semantic relationships. Conversely, low homogeneity implies that the grouping mixes unrelated binaries, signaling that the similarity representation may be noisy or insufficient for distinguishing functionally similar code.

\section{Results}
\label{sec:result}

\subsection{Autoencoder}

\begin{table}[h]
\centering
\caption{Label homogeneity (ssdeep~\cite{ssdeep})}
\label{tab:label_homogeneity}
\begin{tabular}{@{}lcccccc@{}}
\toprule
\multirow{2}{*}{\textbf{K}} & \multicolumn{2}{c}{\textbf{Benign}} & \multicolumn{2}{c}{\textbf{Malicious}} & \multicolumn{2}{c}{\textbf{All}} \\ \cmidrule(lr){2-3} \cmidrule(lr){4-5} \cmidrule(l){6-7} 
 & \textbf{Mean} & \textbf{Std} & \textbf{Mean} & \textbf{Std} & \textbf{Mean} & \textbf{Std} \\ \midrule
1   & 0.51  & 0.49  & 0.8   & 0.39  & 0.66  & 0.47  \\
10  & 3.55  & 4.34  & 7.31  & 4.16  & 5.43  & 4.65  \\
50  & 12.31 & 19.24 & 32.59 & 22.12 & 22.45 & 23.07 \\
100 & 19.84 & 34.82 & 61.00 & 45.22 & 40.42 & 45.31 \\ \bottomrule
\end{tabular}
\end{table}

The unsupervised Autoencoder model discussed in ~\ref{subsect:autoencoder} for dimensionality reduction for the data set, showed better performance than the ssdeep hashing results for the Top-K similar values. 
Table~\ref{tab:label_homogeneity_autoencoder} shows unsupervised autoencoder model for the Top-K consistently shows above 80\% better similarity as compared to the 40\% for ssdeep in Table~\ref{tab:label_homogeneity} 

\begin{table}[h]
\centering
\caption{Label homogeneity (Autoencoder)}
\label{tab:label_homogeneity_autoencoder}
\begin{tabular}{@{}lcccccc@{}}
\toprule
\multirow{2}{*}{\textbf{K}} & \multicolumn{2}{c}{\textbf{Benign}} & \multicolumn{2}{c}{\textbf{Malicious}} & \multicolumn{2}{c}{\textbf{All}} \\ \cmidrule(lr){2-3} \cmidrule(lr){4-5} \cmidrule(l){6-7} 
 & \textbf{Mean} & \textbf{Std} & \textbf{Mean} & \textbf{Std} & \textbf{Mean} & \textbf{Std} \\ \midrule
1   & 0.89 & 0.32  & 0.90 & 0.31  & 0.89 &	0.31  \\
10  & 8.44 & 2.33  & 7.31  & 8.56 & 8.5 &	2.48  \\
50  & 40.27 & 11.57 & 32.59 & 40.87 & 40.57 &	12.52 \\
100 & 78.69 & 23.12 & 79.69 & 27.18 & 79.19	& 25.24 \\ \bottomrule
\end{tabular}
\end{table}

\subsection{Binary (Malware) Classifiers}

Binary classifier performance is summarized in Table~\ref{tab:binary_results}. XGBoost outperforms the neural network across all metrics, particularly in overall accuracy and F1-score. The extremely high AUC (0.9971) indicates near-perfect separability between benign and malicious samples using static features alone.

These results demonstrate that gradient-boosted trees remain highly competitive for structured PE metadata and are particularly well suited for high-accuracy binary detection tasks.

\begin{table}[t]
\centering
\caption{Malware classification performance}
\label{tab:binary_results}
\begin{tabular}{lccccc}
\toprule
\textbf{Model} &
\textbf{Accuracy} &
\textbf{Precision} &
\textbf{Recall} &
\textbf{F1 Score} &
\textbf{AUC} \\
\midrule
XGBoost (Binary) &
\textbf{0.9776} &
\textbf{0.9776} &
\textbf{0.9776} &
\textbf{0.9776} &
\textbf{0.9971} \\
Deep Learning (Binary) &
0.9604 &
0.9553 &
0.9661 &
0.9607 &
0.9926 \\
\bottomrule
\end{tabular}
\end{table}

\subsection{Multi-class (AVClass) classifiers}

\begin{table}[t]
\centering
\caption{Global AVClass classification accuracy metrics}
\label{tab:multiclass_global}
\begin{tabular}{lccc}
\toprule
\textbf{Model} &
\textbf{Accuracy} &
\textbf{Top-5 Accuracy} \\
\midrule
XGBoost (AVClass) &
\textbf{0.9055} &
\textbf{0.9748} \\
Deep Learning (AVClass) &
0.8015 &
0.9244 \\
\bottomrule
\end{tabular}
\end{table}

Global accuracy results for AVClass classification are reported in Table~\ref{tab:multiclass_global}. XGBoost outperforms the deep learning classifier by more than 10 percentage points in overall accuracy. The high Top-5 accuracy (97.48\%) indicates that the correct family label is almost always among the model’s highest-confidence predictions, which is valuable for malware triage workflows where analysts consider multiple candidate families.

\begin{table}
\centering
\caption{AVClass classification performance (macro-averaged)}
\label{tab:multiclass_macro}
\begin{tabular}{lcccc}
\toprule
\textbf{Model} &
\textbf{Precision} &
\textbf{Recall} &
\textbf{F1 } &
\textbf{ROC-AUC (OvR)} \\
\midrule
XGBoost (AVClass) &
\textbf{0.8697} &
\textbf{0.8155} &
\textbf{0.8330} &
\textbf{0.9969} \\
Deep Learning (AVClass) &
0.6655 &
0.7946 &
0.6968 &
0.9916 \\
\bottomrule
\end{tabular}
\end{table}

\begin{table}
\centering
\caption{AVClass classification performance (weighted-averaged)}
\label{tab:multiclass_weighted}
\begin{tabular}{lcccc}
\toprule
\textbf{Model} &
\textbf{Precision} &
\textbf{Recall} &
\textbf{F1} &
\textbf{ROC-AUC (OvR)} \\
\midrule
XGBoost (AVClass) &
\textbf{0.9072} &
\textbf{0.9055} &
\textbf{0.9055} &
\textbf{0.9978} \\
Deep Learning (AVClass) &
0.8015 &
0.8015 &
0.8171 &
0.9926 \\
\bottomrule
\end{tabular}
\end{table}

More detailed macro- and weighted-averaged metrics are shown in Tables~\ref{tab:multiclass_macro} and \ref{tab:multiclass_weighted} respectively. Macro metrics weight each class equally and therefore emphasize performance on less frequent families. The large gap in macro-F1 confirms that XGBoost generalizes more effectively across both dominant and minority families. The smaller gap under weighted averaging reflects improved performance of both models on frequent families, but XGBoost remains superior. Additionally, both models achieve very high one-vs-rest ROC–AUC scores (>0.99), indicating strong probabilistic separation across classes even when discrete classification errors occur.

\subsection{Similarity}

We analyze the AVClass and binary similarity tasks separately since they are structurally very different.

\subsubsection{Multi-class (AVClass) similarity}

\begin{table}
\centering
\caption{Clustering separation and compactness metrics for AVClass classification models}
\label{tab:clustering_sep_avclass}
\begin{tabular}{lccc}
\toprule
\textbf{Model} &
\textbf{Silhouette} $\uparrow$ &
\textbf{Calinski--Harabasz} $\uparrow$ & 
\textbf{Davies--Bouldin} $\downarrow$ \\
\midrule
XGBoost (AVClass) &
-0.4704 &
347.8384 &
22.2092
\\
Deep Learning (AVClass) &
\textbf{0.3287} &
\textbf{990.19} &
\textbf{2.0109} \\
\bottomrule
\end{tabular}
\end{table}

The negative Silhouette score for XGBoost indicates severe overlap between AVClass clusters in Euclidean space. On average, samples are closer to neighboring families than to members of their assigned family. The very high Davies–Bouldin index (22.21) further confirms poor separation and large intra-cluster dispersion. In contrast, the deep learning similarity metrics consistently indicate stronger family-level cohesion and clearer inter-family separation in the neural latent space.

Importantly, this occurs despite XGBoost achieving higher AVClass classification accuracy. This divergence demonstrates that high supervised performance does not guarantee a geometrically well-structured embedding under Euclidean distance.

\subsubsection{Binary (Malware) similarity}

\begin{table}
\centering
\caption{Clustering separation and compactness metrics for malware classification models}
\label{tab:clustering_sep_label}
\begin{tabular}{lccc}
\toprule
\textbf{Model} &
\textbf{Silhouette} $\uparrow$ &
\textbf{Calinski--Harabasz} $\uparrow$ & 
\textbf{Davies--Bouldin} $\downarrow$ \\
\midrule
XGBoost (Binary) &
0.1068 &
5757.2693 &
2.6930 \\
Deep Learning (Binary) &
\textbf{0.3043} &
\textbf{31917.23} &
\textbf{1.3816} \\
\bottomrule
\end{tabular}
\end{table}

In the binary setting, both models produce positive Silhouette scores, indicating meaningful separation between benign and malicious samples. However, the deep learning embeddings again show substantially stronger structure.

Overall, this suggests that tree-based partitions are sufficient to separate two broad classes but do not scale well to fine-grained, high-cardinality family structure.

\subsection{Label Homogeneity@K Evaluation (XGBoost Leaf Similarity)}
\label{subsect:label_homogeneity}

To enable direct comparison with EmberSim \cite{crwdstrkepaper}, we reproduce their Label Homogeneity@K evaluation using XGBoost binary classifier's leaf-index similarity. This metric measures the number of Top-K nearest neighbors that share the same label as the query sample. Higher values indicate stronger class-consistent neighborhoods. A perfect score at K = 1 means the nearest neighbor always shares the same label.

We evaluate two configurations:
\begin{itemize}
    \item XGBoost trained without the AVClass feature
    \item XGBoost trained with the AVClass feature included
\end{itemize}

This distinction is critical because AVClass is not always readily available in operational malware pipelines.

\begin{table}
\centering
\caption{Training Without AVClass Feature}
\label{tab:xgboost_without_avclass}
\begin{tabular}{@{}lcccccc@{}}
\toprule
\multirow{2}{*}{\textbf{K}} & \multicolumn{2}{c}{\textbf{Benign}} & \multicolumn{2}{c}{\textbf{Malicious}} & \multicolumn{2}{c}{\textbf{All}} \\ \cmidrule(lr){2-3} \cmidrule(lr){4-5} \cmidrule(l){6-7} 
 & \textbf{Mean} & \textbf{Std} & \textbf{Mean} & \textbf{Std} & \textbf{Mean} & \textbf{Std} \\ \midrule
1   & 0.89  & 0.31  & 0.89  & 0.31  & 0.89  & 0.31  \\
10  & 8.40  & 2.30  & 8.06  & 2.68  & 8.24  & 2.49  \\
50  & 39.55 & 10.71 & 36.22 & 12.85 & 38.01 & 11.87 \\
100 & 75.06 & 20.10 & 67.78 & 24.25 & 71.69 & 22.42 \\ \bottomrule
\end{tabular}
\end{table}

\begin{table}
\centering
\caption{Training With AVClass Feature}
\label{tab:xgboost_with_avclass}
\begin{tabular}{@{}lcccccc@{}}
\toprule
\multirow{2}{*}{\textbf{K}} & \multicolumn{2}{c}{\textbf{Benign}} & \multicolumn{2}{c}{\textbf{Malicious}} & \multicolumn{2}{c}{\textbf{All}} \\ \cmidrule(lr){2-3} \cmidrule(lr){4-5} \cmidrule(l){6-7} 
 & \textbf{Mean} & \textbf{Std} & \textbf{Mean} & \textbf{Std} & \textbf{Mean} & \textbf{Std} \\ \midrule
1   & 0.99  & 0.11  & 0.98  & 0.13  & 0.99  & 0.12  \\
10  & 9.86  & 0.41  & 9.78  & 1.41  & 9.82  & 1.01  \\
50  & 49.21 & 1.20  & 48.86 & 7.29  & 49.05 & 5.06  \\
100 & 98.33 & 1.83  & 97.70 & 14.70 & 98.04 & 10.12 \\ \bottomrule
\end{tabular}
\end{table}

Table~\ref{tab:xgboost_without_avclass} shows that when trained without the AVClass feature, nearest-neighbor consistency remains strong at small K, but performance degrades steadily as K increases. K = 100 nearly 30\% of neighbors do not share the same label.

Table~\ref{tab:xgboost_with_avclass} shows that when the AVClass field is explicitly included in the XGBoost training features, homogeneity increases dramatically. These results closely mirror the near-perfect homogeneity reported in the EmberSim paper.

The discrepancy is very likely driven by the fact that AVClass is itself a strong, often direct indicator of maliciousness, which creates a form of supervision shortcut

\subsection{Label Homogeneity@K Evaluation (Deep Learning Malware Classifier)}

In addition to the XGBoost leaf-index embeddings, we evaluate the 128-dimensional embedding layer extracted from the binary deep learning classifier described in Section 4.2. Table~\ref{tab:nn_label_homogeneity} reports the corresponding Label Homogeneity@K results under Euclidean nearest-neighbor retrieval.

\begin{table}
\centering
\caption{Label Homogeneity (Neural Network Embeddings)}
\label{tab:nn_label_homogeneity}
\begin{tabular}{@{}lcccccc@{}}
\toprule
\multirow{2}{*}{\textbf{K}} & \multicolumn{2}{c}{\textbf{Benign}} & \multicolumn{2}{c}{\textbf{Malicious}} & \multicolumn{2}{c}{\textbf{All}} \\ \cmidrule(lr){2-3} \cmidrule(lr){4-5} \cmidrule(l){6-7} 
 & \textbf{Mean} & \textbf{Std} & \textbf{Mean} & \textbf{Std} & \textbf{Mean} & \textbf{Std} \\ \midrule
1   & 0.96  & 0.19  & 0.96  & 0.19  & 0.96  & 0.19  \\
10  & 9.52  & 1.49  & 9.53  & 1.65  & 9.53  & 1.57  \\
50  & 47.35 & 7.48  & 47.35 & 8.41  & 47.35 & 7.96  \\
100 & 94.54 & 15.04 & 94.48 & 16.99 & 94.51 & 16.04 \\ \bottomrule
\end{tabular}
\end{table}

Neural embeddings consistently exhibit stronger neighborhood purity than XGBoost trained without the AVClass feature (Table~\ref{tab:xgboost_without_avclass}), particularly as K increases. At $K = 100$, the neural model achieves 94.51\% homogeneity compared to 71.69\% for XGBoost without AVClass. This result is significant because it demonstrates that the superiority of neural similarity is not an artifact of the chosen distance function. Even when XGBoost is evaluated using its native leaf-overlap similarity and neural embeddings are evaluated using a generic Euclidean metric, the neural model induces more class-consistent neighborhoods.


\section{Conclusion}

In this paper, we present a systematic comparison of classical similarity digests and learning-based embeddings for malware clustering. Our findings lead to the following conclusions:

\textbf{Superiority of Latent Representations:} Unsupervised Autoencoders consistently outperform traditional fuzzy hashes in similarity retrieval tasks, showing over 80\% label homogeneity compared to roughly 40\% for classical methods.

\textbf{Robustness Across Similarity Metrics} Even when evaluated under different similarity functions, neural embeddings demonstrate superior neighborhood structure. While EmberSim-style XGBoost similarity relies on tree-proximity (Hamming-style similarity) that is specifically aligned with the internal structure of gradient-boosted trees, our deep learning embeddings evaluated purely under Euclidean distance produce higher label homogeneity and stronger cluster cohesion.

\textbf{Task-Specific Model Selection:} While XGBoost is the preferred choice for high-accuracy classification, Deep Learning architectures are better suited for generating embeddings that require high clustering density and separation, performing better across all similarity metrics (cluster separation, pairwise similarity and label homogeneity) compared to XGBoost.

\textbf{Hybrid Approaches are Essential:} Given the distinct trade-offs between byte-centric and feature-centric methods, effective malware analysis platforms must integrate complementary techniques to ensure robustness against various obfuscation strategies.
\section{Future Work}

This study establishes a framework for evaluating learning-based similarity fingerprinting techniques using static PE metadata. Several promising directions remain for future exploration.

\subsection{Evaluation on Raw Binary Corpora and Classical Similarity Digests}
A key limitation of this work is that the EMBER dataset provides only pre-extracted static features rather than access to the raw binary samples themselves. As a result, direct comparison with classical similarity digests such as TLSH and ssdeep is not possible, as these techniques operate on raw byte sequences. Additionally, obtaining large-scale malware corpora with redistributable binaries often requires expensive commercial licenses or restrictive data-sharing agreements. Future work will evaluate the proposed similarity framework on datasets that include raw executables, enabling direct comparison between byte-level fuzzy hashing methods and learning-based embedding approaches under identical experimental conditions.

\subsection{Integration of Dynamic Analysis Features}
Static analysis is inherently limited in its ability to capture runtime behavior such as control-flow dynamics, system calls, and network interactions. Future work will explore the integration of dynamic analysis features extracted from frameworks such as \texttt{capa} and \texttt{CAPE}. These tools expose high-level behavioral semantics, including API usage patterns and malware capabilities, which may complement static embeddings and improve similarity detection for heavily obfuscated or packed samples.

\subsection{Scalability and Approximate Similarity Search}
Although this study focuses on embedding quality and clustering behavior, practical malware analysis systems must operate at the scale of millions to billions of samples. Future work will evaluate the scalability of the proposed embedding representations under approximate nearest-neighbor indexing schemes and explore trade-offs between retrieval accuracy, memory footprint, and query latency.

\subsection{Labeled Similarity Datasets and External Validation}
Another significant avenue for future work is the development and evaluation of similarity models using \emph{labeled similarity data}. In practice, ground-truth similarity annotations such as explicit pairwise similarity judgments or curated malware clone sets are difficult to obtain at scale. As a result, the current study relies on \emph{internal clustering validation metrics} (e.g., Silhouette, Davies--Bouldin, and Calinski--Harabasz indices) rather than external similarity labels.

Together, these directions aim to extend the current framework beyond static feature evaluation and toward a comprehensive, scalable, and behavior-aware similarity system suitable for real-world malware triage and threat hunting.

\bibliographystyle{splncs04}
\bibliography{main}
%




\end{document}